\documentclass[a4paper,10pt]{article}

\usepackage{graphics,epsfig}
\usepackage{amsmath}
\usepackage{amsfonts}
\usepackage{amssymb}

\newcommand{\be}{\begin{equation}}
\newcommand{\ee}{\end{equation}}
\newcommand{\ba}{\begin{eqnarray}}
\newcommand{\ea}{\end{eqnarray}}

\begin{document}

\title{An effective QCD Lagrangian  in the presence of an axial chemical potential}

\author{A. A. Andrianov$^{1,2}$, D. Espriu$^{2}$ and X. Planells$^{2}$\\
\small{$^1$ V. A. Fock Department of Theoretical Physics,  Saint-Petersburg State
University,}\\
{\small 198504 St. Petersburg, Russia}\\
{\small $^2$ Departament d'Estructura i Constituents de la Mat\`eria and }\\
{\small Institut de Ci\`encies del Cosmos (ICCUB),
Universitat de Barcelona,}\\
{\small  Mart\'\i \ i Franqu\`es 1, 08028 Barcelona, Spain}}

\date{October 2012}

\maketitle

\begin{abstract}
We consider the low energy realization of QCD in terms of mesons when an axial chemical potential is present;
a situation that may be relevant in heavy ion collisions. We shall
demonstrate that the presence of an axial charge has profound consequences on meson physics.
The most notorious effect is the appearance of an explicit source of parity breaking. The eigenstates of strong
interactions do not have a definite parity and interactions that would otherwise be forbidden
compete with the familiar ones. In this work we focus on scalars and pseudoscalars that are
described by a generalized linear sigma model. We comment briefly on the screening role of
axial vectors in formation of effective axial charge and on the possible experimental relevance of our results,
whose consequences may have been already seen at RHIC.
\end{abstract}

%%%%%%%%%%%%%%%%%%%%%%%%%%%%%%%%%%%%%%%%%%%%%%%
\vspace{-13cm}
\begin{flushright} ICCUB-12-363\\
 UB-ECM-PF-80/12
\end{flushright}  
\vspace{11.5cm}
%%%%%%%%%%%%%%%%%%%%%%%%%%%%%%%%%%%%%%%%%%%%%%%

\section{Introduction}

The possibility that in extreme conditions QCD breaks parity has been in the past years actively investigated.
Indeed invariance under parity is one of the characteristic footprints of strong interactions. Yet there are reasons to
believe that this symmetry may be broken in Nature
at least in two different settings.\bigskip

The first possibility is likely to occur in environments with a large baryonic chemical potential. In this case the
fermionic determinant in the QCD partition function is not anymore positive definite and the studies that indicate that
parity cannot be spontaneously broken \cite{vw} for $\mu_B=0$ simply do not apply at $\mu_B\neq 0$. In fact analytical
studies with effective meson lagrangians realizing all the QCD properties at low energies suggest that there is a definite
window of baryonic chemical potentials leading to a vacuum where parity is spontaneously broken \cite{anesp}. Unfortunately
it is difficult to verify these results numerically using lattice field techniques due to the notorious problems
that finite density numerical simulations present \cite{latt}.\bigskip

A second possibility has been abundantly discussed in recent times in connection with the so-called Chiral
Magnetic Effect \cite{kharzeev}. It is believed that thermal fluctuations may induce topological charge fluctuations
in the gauge configuration \cite{polya,zhitn} $\Delta T_5$ and they are detected on lattices \cite{polikarpov}. This leads to an effective '$\theta$- term' in the effective
action that in turn induces a coupling between
the electric and the magnetic fields, leading to the production of positively and negatively charged particles in opposite
directions. It has been claimed that this signal is observed in the STAR experiment at RHIC \cite{star}, although the issue
still remains controversial. In addition the precise mechanism of creation of a sufficiently intense topological
fluctuation is also unclear at present.
\bigskip

This fluctuation, if extended over the whole fireball,  may live for a
sufficiently long time to generate in practice a metastable state characterized by a topological chemical
potential $\mu_\theta$.  If this is the case, in a finite volume and only for light quarks, the topological
chemical potential can be transformed into a chiral chemical potential $\mu_5$ thanks to the anomalous PCAC equation
\be
\partial^\mu J_{5,\mu} -2 i \bar q \hat m_q \gamma_5 q = \frac{N_f}{2\pi^2} \partial^\mu K_\mu \label{pcac},
\ee
where
\be
K_\mu =  \frac12 \epsilon_{\mu\nu\rho\sigma}\text{Tr}\left(G^\nu\partial^\rho G^\sigma -i\frac23 G^\nu G^\rho G^\sigma\right),
\ee
and
\be
\Delta T_5 = T_5(t_f) - T_5(0) = \frac{1}{4\pi^2}\int^{t_f}_0 dt\int_{\text{vol.}}d^3x \, \partial^\mu K_\mu.
\ee
If $m_q\simeq 0$ and  no zero modes are present due to the finiteness of the volume then one gets,
\be
\frac{d}{dt}  (Q_5^q - 2 N_f T_5)  \simeq  0,
\qquad  Q_5^q =\int_{\text{\small vol.}}d^3x\, \bar q\gamma_0\gamma_5q \label{axcons}.
\ee
This is the physical situation we will consider
in the present work. In any case, it seems interesting to investigate how hadronic physics is modified by
the presence of $\mu_5$.\\

This paper is organized as follows. In Sec. 2 a generalized $\Sigma$ model is presented.
Mass-gap equations for three v.e.v. of neutral scalar fields are derived and solved analytically.
Then we determine the best fit of parameters of the model comparing with the experimental inputs for
scalar and pseudoscalar meson spectral data \cite{PDG}.
In Sec. 3 the axial chemical potential
is introduced and treated as a constant time component of an
isosinglet axial-vector field in the non-strange sector. We obtain the modification of the
 mass-gap equations and find the distortion of $a_0$- and  $\pi$- meson spectra caused by the parity
breaking. In Sec. 4 a more complicated mixing of three meson states, $\sigma$, $\eta$, $\eta'$, is investigated
when the medium has an axial charge. At certain energies some particle states become tachyons (recall
we are in-medium so this actually does not represent a fundamental problem).
In Sec. 5 all decay widths are calculated for both the rest frame and for moving particles
(we note that when axial charge fills the medium, the Lorentz invariance is broken).
In Sec. 6 the problem of isosinglet axial-vector meson condensation and its interference
with the axial chemical potential is examined. Sec. 7 is devoted to our conclusions and outlook.

\section{Generalized $\Sigma$ model}

Our starting point will be the following Lagrangian, invariant under $SU(3)_F$
\begin{align}
\nonumber \mathcal L=&\frac 14 \text{Tr}\left (D_\mu HD^\mu H^\dagger\right )+
\frac b2 \text{Tr}\left [M(H+H^\dagger)\right ]+
\frac{M^2}2 \text{Tr}\left (H H^\dagger\right ) -
\frac{\lambda_1}2 \text{Tr}\left [(HH^\dagger)^2\right ]-\frac{\lambda_2}4 \left [\text{Tr}\left (H H^\dagger\right )\right ]^2\\
&+\frac c2 (\text{det}H + \text{det}H^\dagger) +
\frac{d_1}2 \text{Tr}\left [M(H H^\dagger H+H^\dagger H H^\dagger)\right ]+
\frac{d_2}2 \text{Tr}\left [M(H+H^\dagger)\right ] \text{Tr}\left (H H^\dagger\right )\label{lageff}
\end{align}
where $H=\xi \Sigma\xi$, $\xi=\exp\left(i\frac{\Phi}{2f}\right )$, $\Phi=\lambda^a\phi^a$
and $\Sigma=\lambda^b\sigma^b$. This model can be confronted to similar models in \cite{boguta,shechter,rischke} with the important difference (see below) in the trilinear vertices with couplings $d_1, d_2$. The neutral v.e.v. of the scalars are defined as $v_i=\langle\Sigma_{ii}\rangle$
where $i=u,d,s$, and satisfy the following gap equations:
\begin{equation}
M^2v_i-2\lambda_1 v_i^3-\lambda_2\vec v^2v_i+c\frac{v_uv_dv_s}{v_i}=0.
\end{equation}
For further purposes we need the non-strange meson sector and $\eta_s$.
In terms of v.e.v. and physical scalar and pseudoscalar states, the parametrization used here is
\begin{equation}
\Phi=\begin{pmatrix}
\eta_q+\pi^0 & \sqrt{2}\pi^+ & 0\\
\sqrt 2 \pi^- & \eta_q-\pi^0 & 0\\
0 & 0 & \sqrt 2\eta_s
\end{pmatrix}, \qquad \Sigma=\begin{pmatrix}
v_u+\sigma+a_0^0 & \sqrt 2 a_0^+ & 0\\
\sqrt 2 a_0^- & v_d+\sigma-a_0^0 & 0\\
0 & 0 & v_s
\end{pmatrix}.
\end{equation}
The mixing among $\eta$'s is defined via the $\psi$ angle
\begin{equation}
\begin{pmatrix}
\eta_q\\
\eta_s
\end{pmatrix}=\begin{pmatrix}
\cos\psi & \sin\psi\\
-\sin\psi & \cos\psi
\end{pmatrix}\begin{pmatrix}
\eta\\
\eta'
\end{pmatrix}.
\end{equation}

As stated in the introduction, an axial charge can only be effectively generated for nearly massless quarks. Therefore
we exclude kaons and their scalar partners $\kappa$ from the analysis.
Let us take for the time being
$\mu_5=0$ and  assume $v_u=v_d=v_s=v_0\equiv f_\pi\approx 92$ MeV because we only consider the effect of masses
perturbatively.
As a function of the Lagrangian parameters, the main physical magnitudes derived from this model are
\begin{align}
\nonumber v_0&=\frac{c\pm\sqrt{c^2+4M^2(2\lambda_1+3\lambda_2)}}{2(2\lambda_1+3\lambda_2)},
\qquad m^2_\pi= \frac 2{v_0}(b+(d_1+3d_2)v_0^2)m,\\
\nonumber m^2_a&=2(-M^2+ 3(2\lambda_1+\lambda_2)v_0^2+cv_0-2(3d_1+2d_2)mv_0-2d_2m_sv_0),\\
\nonumber m^2_\sigma&= 2(-M^2+(6\lambda_1+7\lambda_2)v_0^2-cv_0-6(d_1+2d_2)mv_0-2d_2 m_sv_0),\\
\nonumber m^2_{\eta,\eta'}&=\frac{m^2_\pi}{2m}(m+m_s)+3cv_0\mp\sqrt{8c^2 v_0^2+\left [\frac{m^2_\pi}{2m}(m-m_s)+cv_0\right ]^2},\\
\nonumber \Gamma^2_a&=\frac{(-4m^2_\eta m^2_\pi+(-m^2_a+m^2_\eta+m^2_\pi)^2)(m^2_a-m^2_\eta +4d_1mv_0)^4}{(48 m^3_a \pi v_0^2)^2},\\
\nonumber \Gamma^2_\sigma&=\frac{9(m^2_\sigma-4 m^2_\pi)(m^2_\sigma-m^2_\pi+4(d_1+2d_2)mv_0)^4}{(32m^2_\sigma \pi v_0^2)^2},\\
\sin(2\psi)&=\frac{4\sqrt{2}cv_0}{m_{\eta'}^2-m_\eta^2},
\end{align}
with $m/m_s=(m_u+m_d)/(2m_s)\simeq 1/25$. The angle $\psi$ is not really necessary for our subsequent discussion
 but it will be eventually useful in the computation of Dalitz decays and we can use it as a test of the model.
$v_0$ is found via gap equations. These equations are inserted in MINUIT in order to
find the minimum of the $\chi^2$ estimator using the following experimental numbers (in MeV):
\begin{gather}
\nonumber v_0^{\text{exp}}=92\pm 5, \quad m_\pi^{\text{exp}}=137\pm 5, \quad m_a^{\text{exp}}=980\pm 50,\\
\nonumber m_\sigma^{\text{exp}}=600\pm 120, \quad m_\eta^{\text{exp}}=548\pm 50, \quad m_{\eta'}^{\text{exp}}=958\pm 100,\\
\Gamma_a^{\text{exp}}=60\pm 30, \quad \Gamma_\sigma^{\text{exp}}=600\pm 120.
\end{gather}
The $\sigma$ mass is assumed to be relatively large as we are not considering any glueballs, so this $\sigma$
is a sort of average of the real $\sigma$ and other $0^+$ light states. We have assigned generous error bars to include the uncertainties in the model itself. In the minimization process
there are several control variables:
\begin{itemize}
\item The value of the potential in the minimum has to be $V(v_0)<0$ since $v=0$ is an extremal point with $V(v=0)=0$
but in the case the latter is a minimum, the true vacuum has to have a lower energy.
Also, there is a control of the third extremal, which has to be higher than the true minimum.
\be
V(v_0)=\frac 14v_0^2(-6M^2+3(2\lambda_1+3\lambda_2)v_0^2-4cv_0)\nonumber
\ee

\item The Hessian matrix has  degenerate eigenvalues and there are  only two different eigenvalues whose positivity
should be provided
\be
(V'')_1(v_0)=-M^2+3(2\lambda_1+\lambda_2)v_0^2+cv_0,\nonumber
\ee
\be
(V'')_2(v_0)=-M^2+3(2\lambda_1+3\lambda_2)v_0^2 -2cv_0\nonumber.
\ee
\end{itemize}
The final result of the minimization process is given in the following table:
%\bigskip
\begin{center}
\begin{tabular}{||c|c|c|c||}
\hline\hline
Magnitude & \texttt{MINUIT} value (MeV) & Experimental value (MeV) & Error\\\hline
$v_0$ & 92.00 & 92 & $-3.52\times 10^{-7}$\\
$m_\pi$ & 137.84 & 137 & $6.10\times 10^{-3}$\\
$m_a$  & 980.00 & 980 & $-1.26\times 10^{-6}$\\
$m_\sigma$ & 599.99 & 600 & $-1.66\times 10^{-5}$\\
$m_\eta$ & 497.78 & 548 & $-9.16\times 10^{-2}$\\
$m_{\eta'}$ & 968.20 & 958 & $1.06\times 10^{-2}$\\
$\Gamma_a$ & 60.00 & 60 & $2.04\times 10^{-5}$\\
$\Gamma_\sigma$ & 600.00 & 600 & $6.81\times 10^{-6}$ \\\hline\hline
\end{tabular}
\end{center}
All the requirements concerning the control parameters are satisfied at this global solution. The fit makes
the cubic (in $H$) terms in (\ref{lageff}) actually more relevant than the linear one.
As a last point, the $\psi$ angle is treated as a prediction. Experimentally \cite{PDG},
$\psi\simeq -18^{\circ}+\arctan\sqrt 2\simeq -18^{\circ}+54.7^{\circ}\approx 36.7^{\circ}$,
 while our result is $\psi_{\texttt{MINUIT}}\approx 35.46^{\circ}$, in excellent agreement.

\section{Introducing the axial chemical potential}

In order to introduce the axial chemical potential we have to recall that, just as the ordinary baryonic potential
is introduced as the zero-th component of a vector field, the axial chemical potential $\mu_5$ can be implemented
as the time component of an axial-vector field.
We follow the arguments for strange quark suppression \cite{aaep} of parity breaking
effects in fireballs created in heavy ion collisions. These arguments are based on Eq. \eqref{pcac}
where due to the unavoidable left-right oscillations the mean value
of strange quark axial charge is washed out as the strange quark mass is comparable with
decay width of fireballs. Accordingly we will use
axial chemical potential in the non-strange sector only.\\

\begin{figure}[h!]
\centering
\includegraphics[scale=0.3]{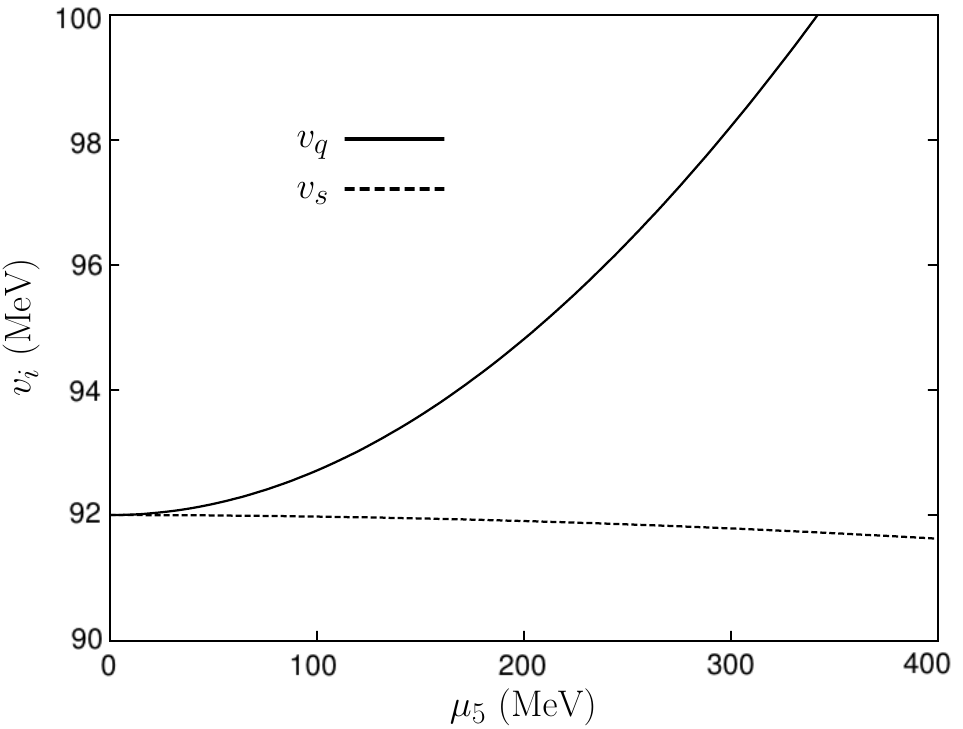}
\vspace{-0.5em}
\caption{$v_q$ and $v_s$ dependence on $\mu_5$.}\label{vqvs}
\vspace{-1em}
\end{figure}

At the level of the meson Lagrangian (\ref{lageff}) it will appear
through the action of the covariant derivative
\be\label{covderiv}
\partial_\mu \to
D_\mu=\partial_\mu -i\{\textbf{I}_q\mu_5\delta_{\mu 0},\cdot\}=\partial_\mu -2i\textbf{I}_q \mu_5\delta_{\mu 0}.
\ee
An extra piece which is proportional to $\mu_5$ appears in the $P$-even Lagrangian
\begin{equation}
\Delta \mathcal L=\frac i2\mu_5 \text{Tr}\left [\textbf{I}_q\left (\partial_0 HH^\dagger-H\partial_0 H^\dagger\right )\right ]+\mu_5^2\text{Tr}\left (\textbf{I}_q HH^\dagger\right ). \label{chempot}
\end{equation}
For non-vanishing $\mu_5$, we will assume isospin symmetry and thus, we impose to our solutions
to have $v_u=v_d=v_q\neq v_s$. The corresponding gap equations are
\be
M^2-2(\lambda_1+\lambda_2)v_q^2-\lambda_2 v_s^2+cv_s +2\mu_5^2 =0,
\ee
\be
v_s(M^2 -2\lambda_2v_q^2-(2\lambda_1+\lambda_2)v_s^2)+cv_q^2=0.
\ee
The correct solution is taken imposing the correct limit $v_q,v_s\to v_0$ when $\mu_5\to 0$ (see Figure \ref{vqvs} for the
$\mu_5$ evolution of the solution).\\

It should be clear that the inclusion of $\mu_5$ leads automatically to a source of parity violation in the
low-energy effective theory. The consequences are far reaching; parity ceases to be a guidance for allowing/forbidding
strong interaction processes, and states that have opposite parities, but otherwise equal quantum numbers, mix.

\bigskip

As an example of such a mixing let us consider the two isotriplets of opposite parity  $\pi$ and $a_0$. After
normalization of the states, let us consider
the piece of the effective Lagrangian that is bilinear in the  $\pi$ and $a_0$ fields
\begin{equation}
\mathcal L=\frac 12 (\partial a_0)^2+\frac 12 (\partial \pi)^2-\frac12 m_1^2 a_0^2-\frac12 m_2^2 \pi^2-4\mu_5 a_0 \dot{\pi},
\end{equation}
where
\begin{align}
\nonumber m_1^2&=-2(M^2-2(3\lambda_1+\lambda_2)v_q^2-\lambda_2 v_s^2-cv_s+2(3d_1+2d_2)mv_q+2d_2m_sv_s+2\mu_5^2),\\
m_2^2&=\frac{2m}{v_q}\left [b+(d_1+2d_2)v_q^2+d_2v_s^2\right ].
\end{align}
Notice that the resulting Lagrangian is not Lorentz invariant, which is obvious from \eqref{covderiv}. We will perform
a diagonalization in momentum space, so the Lagrangian operator is written as
\begin{equation}
\mathcal L=-\frac 12 \begin{pmatrix}
a_0^*(k) & \pi^*(k)
\end{pmatrix}\begin{pmatrix}
-k^2+m_1^2 & 4i\mu_5 k_0\\
-4i\mu_5 k_0 & -k^2+m_2^2
\end{pmatrix}\begin{pmatrix}
a_0(k)\\
\pi(k)
\end{pmatrix}.
\end{equation}
Recall that fields in the momentum representation satisfy $A^*(k)=A(-k)$. Note also the fact that
the mixing term has been rewritten as $-4\mu_5 a_0 \dot{\pi}=-2\mu_5(a_0 \dot{\pi}-\dot{a}_0 \pi)$
in order the matrix to be hermitian. The eigenvalues are $\frac 12(k^2-m_{\text{eff}}^2)$,
where the (energy dependent) effective masses are
\begin{equation}\label{m2effpi}
m^2_{\text{eff}\ \pm}(k_0)=\frac12 \left [m_1^2+m_2^2 \pm \sqrt{(m_1^2-m_2^2)^2+(8\mu_5 k_0)^2}\right ].
\end{equation}
The eigenstates are defined as
\begin{gather}
a_0=\sum_j C_{aj}X_j,\quad \pi=\sum_j C_{\pi j}X_j, \qquad C_{a1}=iC_{\pi 2}=C_+, \quad C_{a2}=-iC_{\pi 1}=-C_-,
\end{gather}
with
\be
C_\pm=\frac 1{\sqrt 2}\sqrt{1\pm\frac{m_1^2-m_2^2}{\sqrt{(m_1^2-m_2^2)^2+(8\mu_5 k_0)^2}}}.
\ee
We can also use the notation $X_1,X_2\equiv \tilde a,\tilde\pi$, indicating that $X_1$ (resp. $X_2$) is the state that when
$\mu_5=0$ goes over to $a_0$ (resp. $\pi$).\\

\begin{figure}[h!]
\centering
\includegraphics[scale=0.3]{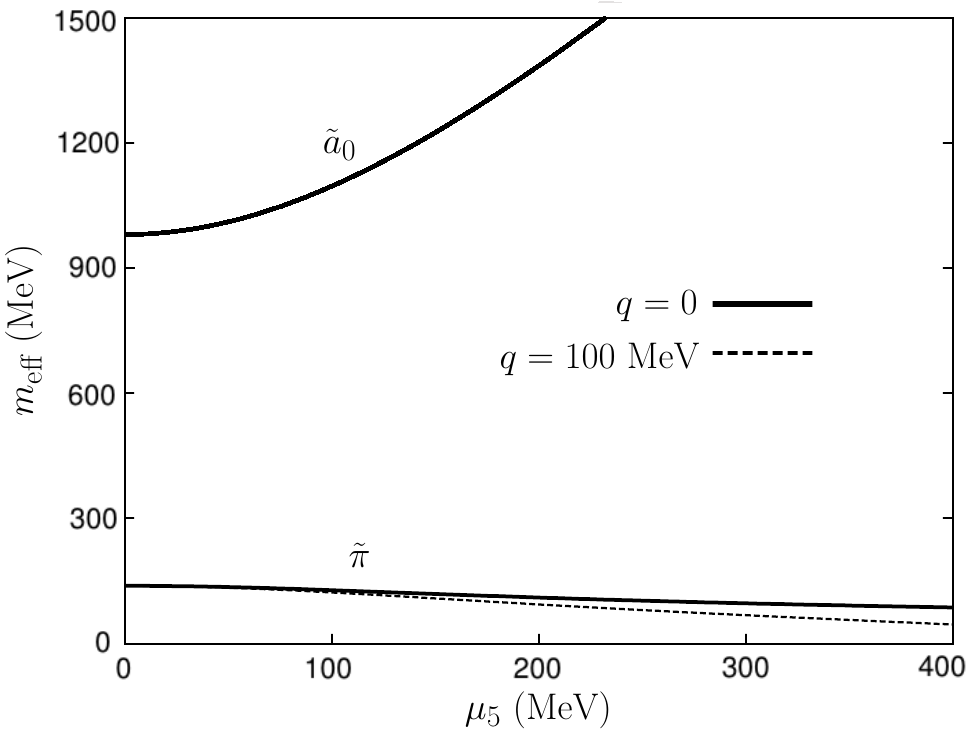}\includegraphics[scale=0.3]{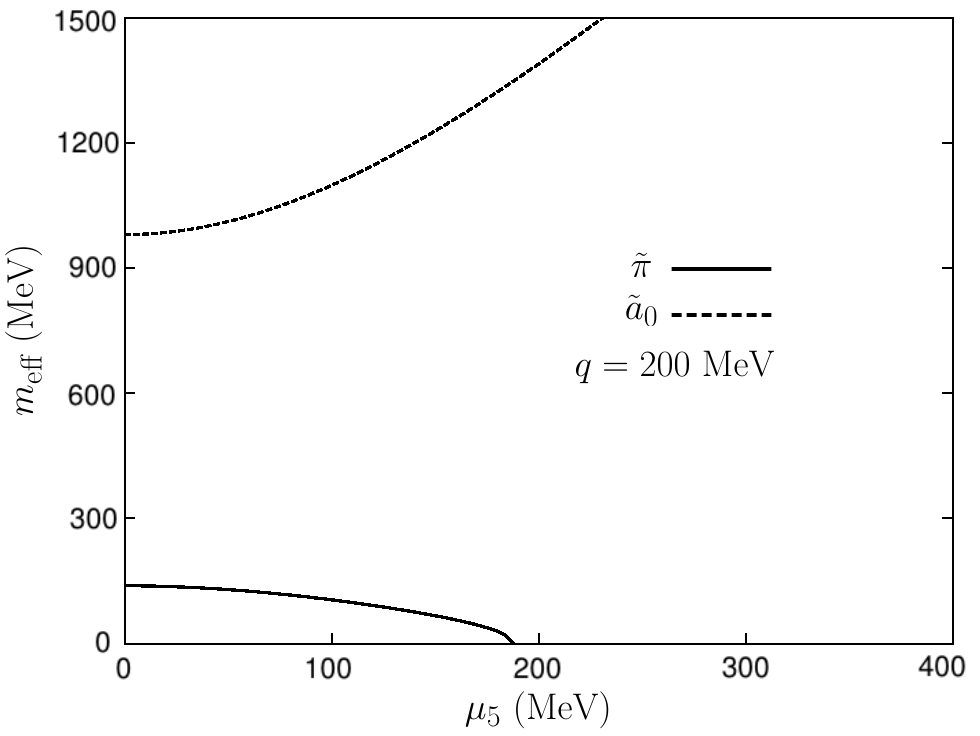}
\vspace{-2em}
\caption{Effective mass dependence on $\mu_5$ for $\tilde\pi$ and $\tilde{a}_0$. Left panel: comparison of masses at rest and at low momentum $q=100$ MeV. Right panel: masses at $q=200$ MeV, where the $\tilde\pi$ mass goes tachyonic, as discussed in the text. For such momenta, the variations in $\tilde a_0$ are almost invisible and only slightly visible for large values of $\mu_5$ for $\tilde \pi$.}\label{meffpi}
\vspace{-1em}
\end{figure}

In Fig. \ref{meffpi}, we present the results for the evolution of $\tilde \pi$ and $\tilde{a}_0$
effective masses with respect to the axial chemical potential $\mu_5$. As stressed, both states
tend to the known physical ones in the limit $\mu_5\to0$. A remarkable feature of this model
is the appearance of tachyonic states at high energies (or momenta). It is evident from Eq. \eqref{m2effpi}
that for energies higher than a critical value $k_0,|\vec k|>m_1m_2/(4\mu_5)\equiv k^c_{\tilde\pi}$,
the square root term dominates, thus leading to a negative squared mass for pions, as shown in the
right panel of Fig. \ref{meffpi}. Such a behaviour does not represent a serious physical obstacle
since it can be checked that the energies are always positive and no vacuum instabilities appear.
On the other hand, the $\tilde{a}_0$ mass shows an important enhancement, but in this model $\mu_5$
has to be understood as a perturbatively small parameter, and very high values are beyond the domain of applicability
of the effective Lagrangian. A better treatment of $\tilde a_0$ would require the inclusion of heavier degrees
of freedom such as $\pi(1300)$ for instance.

\section{Mixing $\eta-\sigma-\eta'$}

A similar analysis applies to the isosinglet case. We shall consider three states here: $\eta$, $\eta^\prime$ and $\sigma$.
As before, the starting point will be the piece of the effective Lagrangian (\ref{lageff}) that after the inclusion
of $\mu_5$ is bilinear in the fields, i.e. the properly normalized kinetic part
\begin{equation}
\mathcal L=\frac 12[(\partial\sigma)^2+(\partial \eta_q)^2+(\partial \eta_s)^2]-\frac 12m_3^2 \sigma^2-\frac 12m_4^2\eta_q^2-\frac 12m_5^2\eta_s^2-4\mu_5\sigma\dot{\eta}_q-2 \sqrt{2}cv_q\eta_q\eta_s.
\end{equation}
The constants appearing in the previous equation are given by
\begin{align}
\nonumber m_3^2&=-2(M^2-6(\lambda_1+\lambda_2)v_q^2-\lambda_2 v_s^2+cv_s+6(d_1+2d_2)mv_q+2d_2m_sv_s +2\mu_5^2),\\
\nonumber m_4^2&=\frac {2m}{v_q}\left [b+(d_1+2d_2)v_q^2+d_2v_s^2\right ]+2cv_s,\\
m_5^2&=\frac{2m_s}{v_s}[b+2d_2v_q^2+(d_1+d_2)v_s^2]+\frac{cv_q^2}{v_s}.
\end{align}
In matrix form
\begin{equation}
\mathcal L=-\frac 12\begin{pmatrix}
\sigma^*(k) & \eta_q^*(k) & \eta_s^*(k)
\end{pmatrix}\begin{pmatrix}
-k^2+m_3^2 & 4i\mu_5 k_0 & 0\\
-4i\mu_5 k_0 & -k^2+m_4^2 & 2 \sqrt{2}cv_q\\
0 & 2 \sqrt{2}cv_q & -k^2+m_5^2
\end{pmatrix}\begin{pmatrix}
\sigma(k) \\
\eta_q(k) \\
\eta_s(k)
\end{pmatrix}.
\end{equation}

The equation for the eigenvalues (effective masses) is now a cubic one and the solution is determined
numerically
\begin{equation}
-8c^2v_q^2(m_{\text{eff}}^2-m_3^2)+(m_{\text{eff}}^2-m_5^2)\left [\left (m_{\text{eff}}^2-\frac 12(m_3^2+m_4^2)\right )^2-\frac 14 (m_3^2-m_4^2)^2-(4\mu_5 k_0)^2\right ]=0.
\end{equation}

As before the eigenstates are defined via
\begin{align}\label{eigens}
\sigma=\sum_j C_{\sigma j}X_j, \qquad \eta_q=\sum_j C_{\eta_q j}X_j, \qquad \eta_s=\sum_j C_{\eta_s j}X_j ,
\end{align}
where
\begin{gather}
\nonumber C_{\sigma j}=\frac{4i\mu_5 k_0 (m_5^2-m_j^2)}{N_j\prod_{k\neq j}(m_j^2-m_k^2)}, \quad C_{\eta_q j}=\frac{(m_5^2-m_j^2)(m_3^2-m_j^2)}{N_j\prod_{k\neq j}(m_j^2-m_k^2)}, \quad C_{\eta_s j}=\frac{-2\sqrt{2}cv_q(m_3^2-m_j^2)}{N_j\prod_{k\neq j}(m_j^2-m_k^2)}\\
\frac 1{N_j}=\sqrt{1+\frac{(4\mu_5 k_0)^2}{(m_j^2-m_3^2)^2}+\frac{8c^2v_q^2}{(m_j^2-m_5^2)^2}}.
\end{gather}
$N_j$ is the proper (eigen)field normalization factor.

\begin{figure}[h!]
\centering
\includegraphics[scale=0.3]{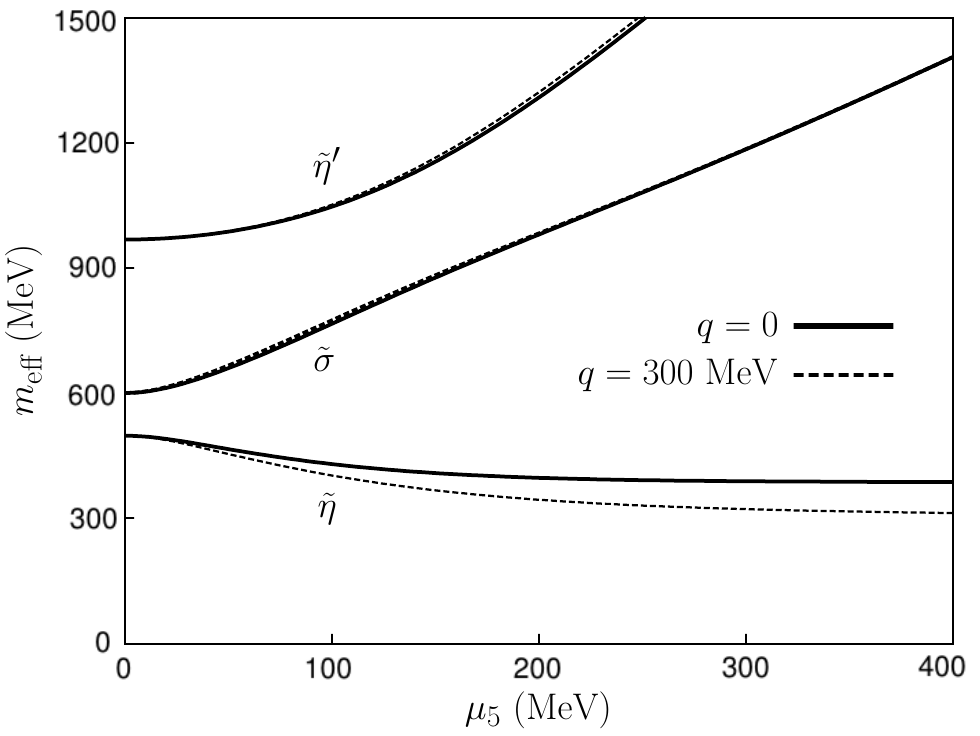}\includegraphics[scale=0.3]{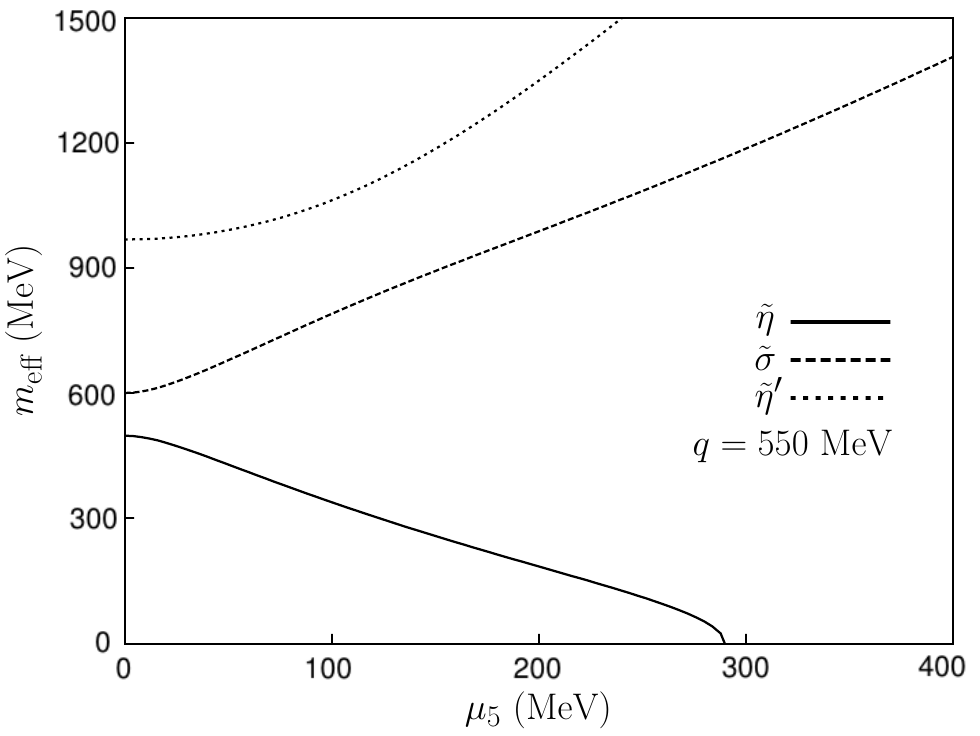}
\vspace{-2em}
\caption{Effective mass dependence on $\mu_5$ for $\tilde\eta$, $\tilde\sigma$ and $\tilde\eta'$. Left panel: comparison of masses at rest and at low momentum $q=300$ MeV. Right panel: masses at $q=550$ MeV, where the $\tilde\eta$ mass goes tachyonic, as discussed in the text. As in the previous example, for this range of momenta, the variations in the heavier degrees of freedom $\tilde \sigma$ and $\tilde\eta'$ are almost invisible and only slightly visible for large values of $\mu_5$ for $\tilde \eta$.}\label{meffeta}
\end{figure}

\begin{figure}[h!]
\centering
\includegraphics[scale=0.3]{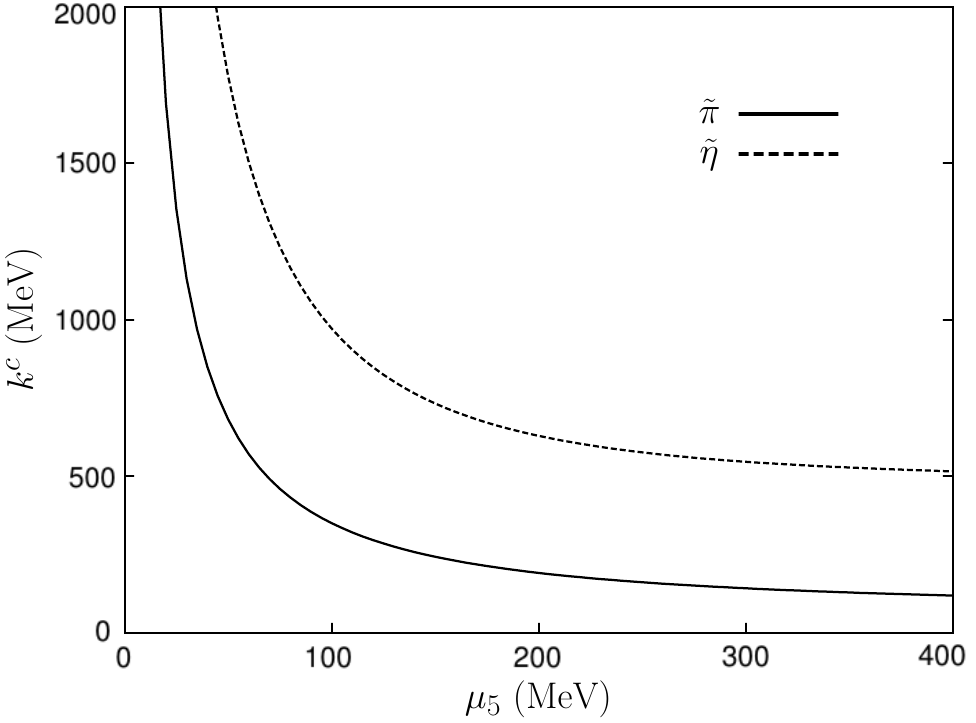}
\vspace{-0.5em}
\caption{$\mu_5$-dependence of the tachyon critical energy for isotriplet $k^c_{\tilde\pi}$ and isosinglet case $k^c_{\tilde\eta}$.}\label{kcrit}
\vspace{-1em}
\end{figure}

The $\mu_5$-dependence of the effective masses is plotted in Fig. \ref{meffeta}. As in the previous example,
the lightest degree of freedom becomes tachyonic for big energies or
momenta $k_0,|\vec k|>k^c_{\tilde\eta}$ with $k^c_{\tilde\eta}\equiv m_3/(4\mu_5m_5)\sqrt{m_4^2m_5^2-8c^2v_q^2}$,
as shown in the right panel of Fig. \ref{meffeta}.
The tachyon critical energy presents a similar behaviour as the one in the triplet case.
In Fig. \ref{kcrit}, both isotriplet $k^c_{\tilde\pi}$ and isosinglet critical energies $k^c_{\tilde\eta}$ are plotted together.

\section{New interactions and decay widths}
After the inclusion of $\mu_5$ the cubic couplings present in the effective Lagrangian \eqref{lageff} are
\begin{align}
\nonumber \mathcal L_{\sigma aa}&=2[(3d_1+2d_2)m-2(3\lambda_1+\lambda_2)v_q]\sigma \vec a^2,\\
\nonumber \mathcal L_{\sigma\pi\pi}&=\frac 1{v_q^2}\left [(\partial\vec\pi)^2v_q-(b+3(d_1+2d_2)v_q^2+d_2v_s^2)m\vec\pi^2\right ]\sigma,\\
\nonumber \mathcal L_{\eta a\pi}&=\frac 2{v_q^2}[\partial \eta_q \vec a\partial \vec\pi v_q-(b+
(3d_1+2d_2)v_q^2+d_2v_s^2)m\eta_q\vec a\vec\pi],\\
\mathcal L_{\sigma a\pi}&=-\frac{4\mu_5}{v_q}\sigma\vec a\dot{\vec{\pi}}, \quad \mathcal L_{\eta aa}=-\frac{2 \mu_5}{v_q}\dot \eta_q \vec a^2,\quad \mathcal L_{\eta\pi\pi} = 0.
\end{align}

As seen in the previous expressions, decays that are normally forbidden on parity conservation grounds are now possible
with a strength proportional to the parity breaking parameter $\mu_5$. However, the
previous interaction terms are not physical because the properly diagonalized states are now $X_i$ rather than the original
fields $\pi$, $a_0$, etc. Going to the physical
basis requires using the diagonalization matrices defined in the previous section.\\

Our ultimate purpose is to check the relevance of dynamically generated parity breaking through topological
charge fluctuations in heavy ion collisions. It is natural then to ask how the previously derived masses and
vertices may influence the physics in the hadronic fireball.\\

It should be clear that the influence may be very important if $\mu_5$ is such that the induced parity
breaking effects are significant. After the initial collision of two heavy ions in a central or quasi-central process
a fireball is formed. This fireball could be described in rather simplistic terms as a hot and dense pion gas.
Pion-pion interaction is dominated by $\sigma$ and $\rho$ -particle exchanges and processes
such as $\eta\to \pi\pi$ or $\eta^\prime \to \pi\pi$ are forbidden. If parity is no longer a restriction, these two
processes, or rather processes such as $X_i \to \tilde\pi \tilde\pi$ ($i=3,4,5$) are for sure relevant and the new
eigenstates produced due to parity breaking could thermalize inside the fireball.\\

Let us now try to be more quantitative. It should be clear from the mass evolution as a function of $\mu_5$ (Figure \ref{meffpi})
that $\tilde\pi$ is the lightest state and it dominates the
partition function in the fireball. As a consequence, to get an estimate of the relevance of these new states
let us compute their width in order to see whether its inverse is comparable to the fireball lifetime. To do so
we need the following $S$-matrix element corresponding to $X_i(q) \to \tilde\pi^+(p) \tilde\pi^-(p')$
\begin{align}
\nonumber i\mathcal M=&4i[(3d_1+2d_2)m-2(3\lambda_1+\lambda_2)v_q]C_{\sigma i} C_{a2}^+C_{a2}^- +\frac{4\mu_5}{v_q}C_{\sigma i}(E_{p'} C_{a2}^+ C_{\pi 2}^- +E_p C_{a2}^- C_{\pi 2}^+) \\
\nonumber &-\frac i{v_q^2}\left [(m_{X_i}^2-m_{\tilde\pi^+}^2-m_{\tilde\pi^-}^2) v_q+2(b+3(d_1+2d_2)v_q^2+d_2v_s^2)m\right ]C_{\sigma i}C_{\pi 2}^+ C_{\pi 2}^-\\
\nonumber &-\frac{4 \mu_5 E_q}{v_q}C_{\eta_q i} C_{a2}^+C_{a2}^- +\frac {i}{v_q}(m_{\tilde\pi^+}^2-m_{\tilde\pi^-}^2)C_{\eta_q i}(C_{a2}^-C_{\pi 2}^+ -C_{a2}^+C_{\pi 2}^-) \\
&-\frac {i}{v_q^2}[2(b+(3d_1+2d_2)v_q^2+d_2v_s^2)m-m_{X_i}^2v_q]C_{\eta_q i}(C_{a2}^+C_{\pi 2}^- +C_{a2}^-C_{\pi 2}^+)
\end{align}
We are dealing with a relativistic non-invariant theory and therefore the widths do not a priori transform
as one would naively think. We shall compute them first at rest, then for different values of the 3-momentum
of the decaying particle.

\subsection{Widths at rest}

\begin{figure}[h!]
\centering
\includegraphics[scale=0.4]{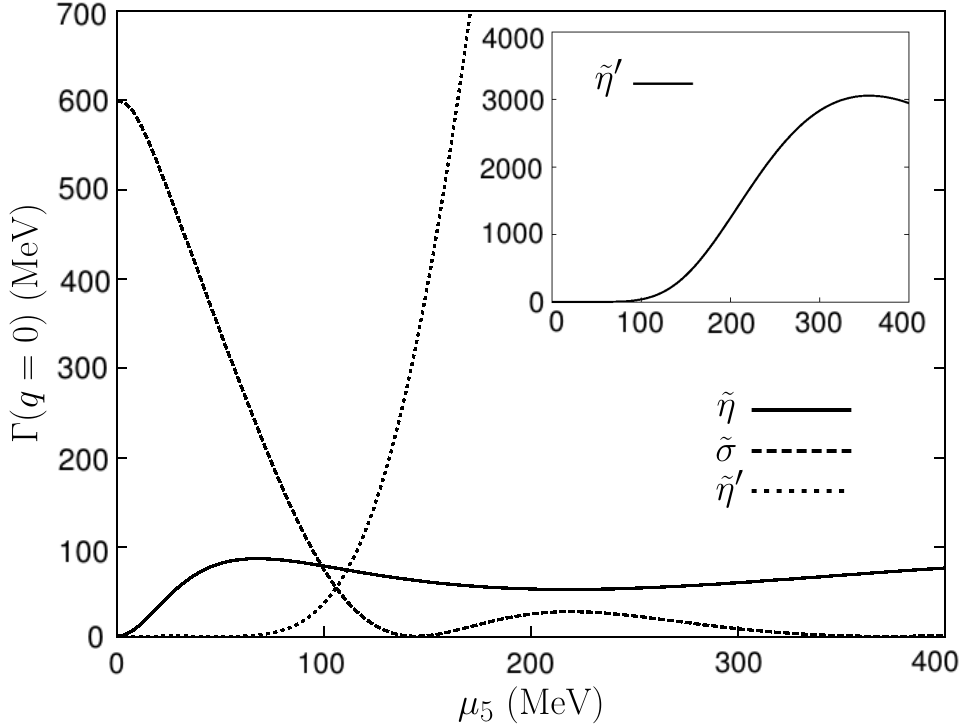}
\vspace{-0.5em}
\caption{$\tilde\eta$, $\tilde\sigma$ and $\tilde\eta'$ widths at rest depending on $\mu_5$. Down to $\mu_5=50$ MeV, $\tilde\eta$ acquires a width of order 60 MeV, with a characteristic mean free path smaller than the typical fireball size of $5\div 10$ fm and hence implying that thermalization may occur in this channel. Nevertheless, $\tilde \sigma$ shows a pronounced fall and beyond $\mu_5=100$ MeV, it becomes a stable channel. Inset: Detail of $\tilde\eta'$ width reaching the GeV scale, a clear violation of unitarity since we don't include heavier degrees of freedom in our model.}\label{width0}
\vspace{-1em}
\end{figure}

The width of $X_i$ is calculated from the amplitude shown before since we don't include further
decaying processes. Recall that all
masses are energy dependent. At the $X_i$ rest frame, $\vec q=\vec 0$ and $E_p=E_{p'}=m_{X_i}(\vec q=\vec 0)/2\equiv m^{X_i}_0/2$.
Here, a momentum-dependent effective mass is taken instead of an energy-dependent one since we assume
the decaying particle to be on-shell so both $m(\vec k)$ and $m(k_0)$ coincide. Thus, the rest width is given by
\begin{equation}
\Gamma_{X_i\to \tilde\pi\tilde\pi}=\frac 32\frac 1{2m^{X_i}_0}|\mathcal M|^2\frac{1}{4\pi}\frac{p_{\tilde\pi}}{m^{X_i}_0\frac{dp_{\tilde\pi}^2}{dE_p^2}}, \qquad \frac{dp_{\tilde\pi}^2}{dE_p^2}=1+\frac{(4\mu_5)^2}{\sqrt{(m_1^2-m_2^2)^2+(8\mu_5 E_p)^2}}
\end{equation}
where $p_{\tilde\pi}(E_p)=\sqrt{E_p^2-m_{\tilde\pi}^2(E_p)}$ and the factor 3/2 accounts for
the decay both to neutral and charged $\tilde\pi$. The results of the widths are shown in Figure \ref{width0}.\\

The new state $\tilde\eta$ exhibits a smooth behaviour with an average value $\sim 60$ MeV,
corresponding to a mean free path $\sim 3$ fm, which is smaller than the typical fireball
size $L_\text{fireball}\sim 5\div 10$ fm. Hence, the thermalization of this channel via
regeneration of $\tilde\pi$ within the gas seems to be possible.\\

Another striking point concerning $\tilde\sigma$ takes place down to $\mu_5\sim 100$ MeV, when
the decay width decreases dramatically leading to scenarios where this state becomes stable.
The visible bumps in these two latter channels seem to reflect the tachyonic nature of the decaying $\tilde\pi$.\\

Finally, we present in the inset of Fig. \ref{width0} the detail of the $\tilde\eta'$ width, that grows
up to the GeV scale, showing clear violations of unitarity. As in the case of $\tilde a_0$, more
hadronic degrees of freedom are needed to obtain a reliable result, such as $f_0(980)$, etc.

\subsection{Decay widths of moving particles}

Next, let us compute the width when the decaying particle is not at rest. As explained before in a non-relativistic
theory this cannot be obtained from the one at rest by simply taking into account the time dilatation effect. Then
\begin{equation}
\Gamma_{X\to \tilde\pi\tilde\pi}=\frac 32\frac 1{2E_q}\frac{1}{8\pi q}\int \frac{|\mathcal M|^2pdp }{E_p\frac{dp_{\tilde\pi}^2}{dE_p^2}}\Theta(1-|\cos\theta|)\Theta(E_{p'})
\end{equation}
where the Heaviside step functions are introduced to make sure that both $\cos\theta$ and $E_{p'}$
take physical values in the numerical calculation. Of course, the limit $q\to 0$ coincides with the
calculation at rest performed before.%\\

\begin{figure}[h!]
\centering
\includegraphics[scale=0.3]{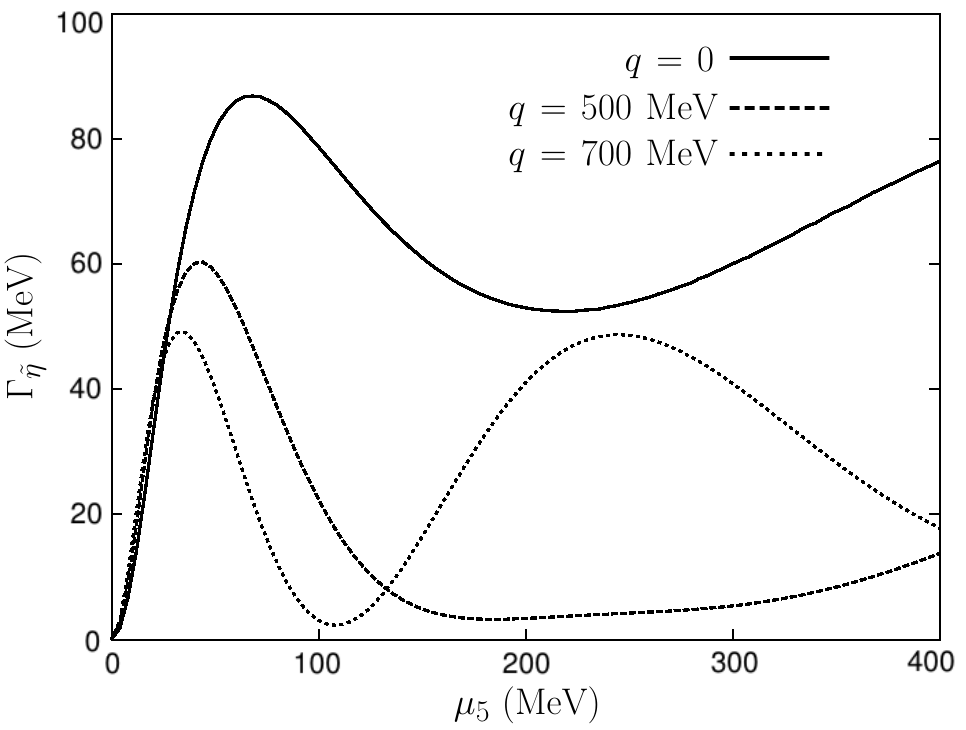}\includegraphics[scale=0.3]{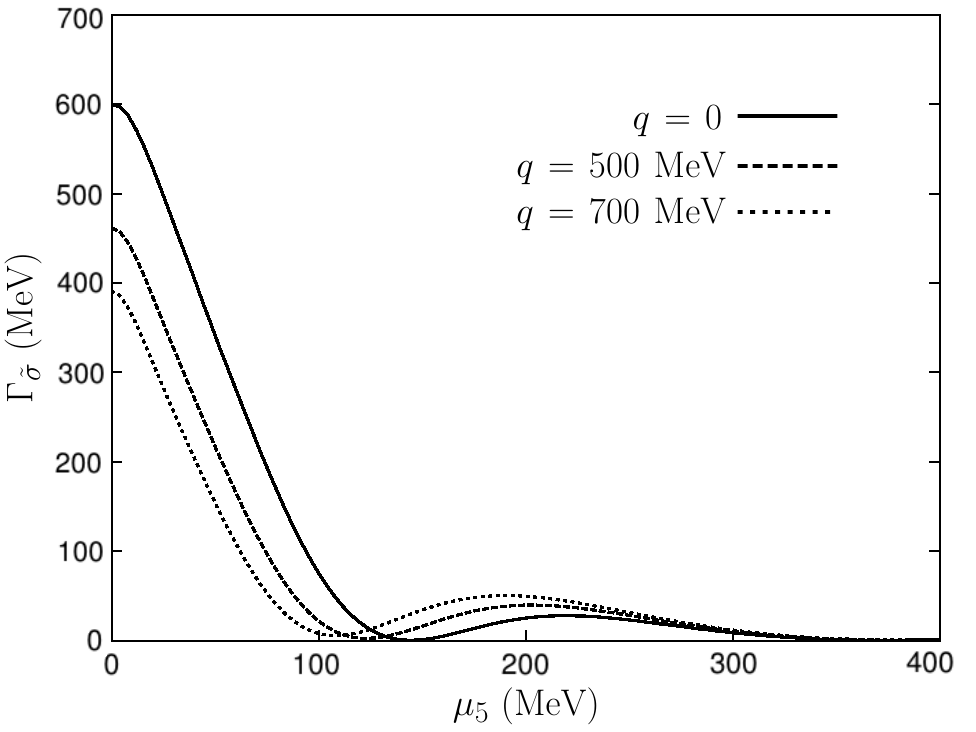}
\vspace{-2em}
\caption{$\tilde\eta$ (left) and $\tilde\sigma$ (right) widths depending on $\mu_5$ for different values of the incoming 3-momentum: $q=0,500,700$ MeV. The first plot shows a non-trivial dependence on $q$ (see text) while the second one shows a fall that is mainly due to the Lorentz factor.}\label{widthq}
%\vspace{-1em}
\end{figure}

In the $\tilde\eta$ channel (see the left section of Fig. \ref{widthq}), one may observe small
variations at low 3-momenta with respect to the width at rest, namely, two initial bumps at $\mu_5\sim 80$ MeV
and 550 MeV (the latter being beyond the plot range) slowly separate
as one increases $q$. However, the two-dimensional representation $\Gamma_{\tilde\eta}(\mu_5,q)$ exhibits a
saddle point around $\mu_5^*\sim 240$ MeV and $q^*\sim 500$ MeV, and in consequence, for large 3-momenta,
a third intermediate bump appears opening the possibility of creating two different tachyons at the same time.
The latter maximum grows fast as one increases $q$ and becomes the global one when the 3-momentum goes
beyond $q\gtrsim 700$ MeV.

On the other hand, in the $\tilde\sigma$ and $\tilde\eta'$ channels no huge differences arise when
boosting the decaying particle. In the right panel of Fig. \ref{widthq}, we show the $\tilde\sigma$ width for
different values of $q$ and the most salient behaviour is, as in the previous case, the separation of the
two minima as one increases $q$.

\section{Axial-vector meson condensation}
The introduction of axial chemical potential into the quark-meson model interferes with
the flavor-singlet axial-vector channel as this potential is just a time component of axial-vector field,
$\Delta {\cal L}_{\mu_5} = \mu_5 \bar q \gamma_0 \gamma_5 \textbf{I}_q q$. Therefore if one includes
the coupling of the singlet axial-vector quark current with the corresponding meson field $h_\mu$ one expects mixing and
renormalization of the bare axial chemical potential due to condensation of the time components of the axial-vector
fields $h^\mu \simeq \langle h^0\rangle \delta^{0\mu}$. This phenomenon is in full analogy to the condensation of
the time component of the $\omega$ meson field when a baryon chemical potential enters the Lagrangian \cite{walec}
which is quite important to understand the repulsive nuclear forces in this channel.\\

Let us elucidate this phenomenon in more details. The relevant Lagrangian for axial-vector mesons reads
\begin{equation}
\Delta\mathcal L=-\frac 14 h_{\mu\nu}h^{\mu\nu}+\frac 12m_h^2h_\mu h^\mu+ \bar q\gamma_\mu\gamma_5 (g_hh^\mu+\delta^{\mu0}\mu_5)\textbf{I}_q q,
\end{equation}
where $h$ stands for the axial-vector meson $h_1(1170)$ \cite{PDG} singlet in flavor and $g_h$ denotes
its coupling to the quark current. We assume the condensation of $h^\mu \simeq \langle h^0\rangle \delta^{0\mu}$
\begin{equation}
 \frac{\delta\Delta\mathcal L}{\delta h_0} = m^2_h \langle h_0\rangle + g_h \langle \bar q\gamma_0\gamma_5\textbf{I}_q q\rangle = 0,
\end{equation}
 so that the effective chemical potential $\bar\mu_5\equiv\mu_5+g_h\langle h_0\rangle$ arises and
determines the effective non-strange axial-charge density $\rho_5 = \langle \bar q\gamma_0\gamma_5 \textbf{I}_q q\rangle$
\begin{equation}\label{deltav}
\rho_5(\bar\mu_5)=\frac{\mu_5-\bar\mu_5}{G_h} = \frac{\delta\Delta\mathcal L}{\delta\bar\mu_5},\quad \Delta V=-\frac 12m_h^2\langle h_0\rangle^2=-\frac 12\frac{(\bar\mu_5-\mu_5)^2}{G_h},\qquad
\end{equation}
where $G_h=g_h^2/m_h^2$. Therefrom we can see that the axial charge density is directly related
to the axial-vector condensate.  After including $\Delta V$ (Eq. \eqref{deltav}) in Eq. \eqref{chempot}
the stationary point equation can be derived
\begin{equation}
\frac{\delta \mathcal L}{\delta \bar\mu_5}=\frac{\delta}{\delta \bar\mu_5}\left [2\bar\mu_5^2 v_q^2+\frac 12\frac{(\bar\mu_5-\mu_5)^2}{G_h}\right ]=0
\end{equation}
that allows to relate the bare and effective axial chemical potentials
\begin{equation}
\bar\mu_5\left [1+4G_h v_q^2(\bar\mu_5)\right ]=\mu_5. \label{renchem}
\end{equation}
We stress that in the mass-gap equations for $v_q, v_s$ the effective axial chemical potential
 $\bar\mu_5$ must be used. The relation \eqref{renchem} is smooth against the decoupling of axial-vector mesons $g_h \to 0$.
It determines unambiguously the axial charge density,
$\rho_5(\bar\mu_5)=4\bar\mu_5 v_q^2(\bar\mu_5)$, which exhibits, in general, lower values when affected
 by axial meson forces, $|\bar\mu_5| < |\mu_5|$.

\section{Conclusions and outlook}

Perhaps the main conclusion of the study of meson physics in an environment endowed with a net axial charge is that
is full of surprises. The axial chemical potential provides a source of parity violation. This makes states of different
intrinsic parities mix and allows for `exotic' processes in hadronic physics. The presence of the axial charge also leads unavoidably to a breaking of Lorentz invariance. The effective in-medium masses are
energy dependent and meson physics is frame-dependent, with the natural consequence that widths or decay rates
do depend non-trivially on the momentum of the decaying particle.
\bigskip

We have assumed that the parity breaking parameter -the axial chemical potential- is an $SU(2)$ singlet. This is natural if the
mechanism to generate $\mu_5$ is via topological charge fluctuations as advocated by some \cite{kharzeev,zhitn}. Note, however, that it is not
an $SU(3)$ singlet, as the topological charge fluctuations are not transmitted to the strange sector (and even less to the
eventual charm sector).\\

It is natural to ask whether such a mixing of states of different parities occurs in the vector/axial-vector sector. There
 are many models dealing with vector particles phenomenologically. If we assume that the vector mesons appear as part of a
covariant derivative (as postulated e.g. in hidden symmetry models \cite{bando}), no mixing term can be generated by operators of
dimension 4 if $\mu_5$ is an isosinglet. However, such a mixing is not forbidden on (global) symmetry grounds if $\mu_5$ is present, appearing as the time component of an axial-vector field (see e.g. \cite{rischke}). This means that this coupling is very much model dependent and, unfortunately, not much phenomenological information is present. It is however an interesting point we plan to analyze in the future.\\

However parity breaking via a topological charge or axial chemical potential influences vector mesons (and eventually
photons too) in a different way discussed in detail in  \cite{aaep}. Their polarizations are severely distorted and the breaking
of Lorentz invariance, together with parity, reflects itself in different polarizations acquiring different effective masses, which could hopefully be experimentally measured. This
issue has not been discussed here.\\

As argued in the introduction, many authors support the idea that topological charge fluctuations may lead to visible effects
via the Chiral Magnetic Effect. If this is so, not only peripheral collisions (where the Chiral Magnetic Effect is present) will
show traces of parity breaking. We have argued elsewhere \cite{aaep, avep} that parity breaking induced from
topological charge fluctuation will lead to possibly measurable effects in central collisions, in the dilepton spectrum from $\rho$ and $\omega$ decays.\\

What we have seen in the work presented here is that the physics of spin zero resonances is also strongly affected by the presence
of an axial chemical potential. We have given convincing arguments that, if $\mu_5\neq 0$ the pion gas in the fireball forming after
a central heavy ion collision may actually not be made of the usual pions, but instead of some states of non-defined parity and
energy-dependent effective mass. In addition all the lightest spin zero states have the same properties and perhaps more
importantly, they are all in thermal equilibrium with the `pion' gas, as indicated by the characteristic large widths, completely
different from the ones in vacuum. These particles have Dalitz decays that are therefore completely distorted with respect to the
$\mu_5=0$ case usually considered. This phenomenon may help in explaining the anomalous dilepton yield enhancement
observed \cite{reviews} for low dilepton invariant masses in heavy ion collisions.

\section*{Acknowledgements}

We acknowledge the financial support from projects FPA2010-20807, 2009SGR502, CPAN (Consolider CSD2007-00042).
A. A. Andrianov is also supported by Grant RFBR 10-02-00881-a.
X. Planells acknowledges the support from Grant FPU AP2009-1855.

\end{document}